\documentclass[%
preprint,
 amsmath,amssymb,
 aps, physrev,
]{revtex4-2}

\usepackage{graphicx}
\usepackage{dcolumn}
\usepackage{bm}
\usepackage{natbib}

\usepackage{graphicx}
\usepackage{tikz}
\usetikzlibrary{arrows.meta, decorations.markings, patterns}

\begin{document}

\preprint{APS/123-QED}

\title{\textbf{Ground effect on Undulation and pumping near surfaces } 
}%

\author{Sunghwan Jung}
 \email{Contact author: sj737@cornell.edu}
\affiliation{%
 Department of Biological and Environmental Engineering, Cornell University, Ithaca, NY 14853, U.S.A.
}%

\date{\today}

\begin{abstract}
Locomotion and fluid pumping near surfaces are ubiquitous in nature, ranging from the slow crawling of snails to the rapid flight of bats. This study categorizes these behaviors based on the Undulation number ($\text{Un}$) and Reynolds number ($Re$). We contrast low $Re$ undulatory propulsion ($\text{Un} > 1$), exemplified by freshwater snails, with high $Re$ flapping propulsion ($\text{Un} < 1$), seen in bats and bees. For snails, we derive lubrication models showing that pumping and swimming speeds scale with $(a/h_0)^2$, a result validated by robotic experiments which also reveal the detrimental effects of surface deformation (high Capillary/Bond ratio). Conversely, for high $Re$ fliers, we examine the ground effect's role in lift enhancement. Biological data from bats (\textit{R. ferrumequinum}) reveal a 2.5-fold increase in lift coefficient during surface-skimming drinking flights, attributed to aerodynamic squeezing effects. Finally, we analyze honeybee fanning, demonstrating how a "jet-vortex" mechanism utilizes ground effect to transport pheromones efficiently against diffusion. These findings provide a unified framework for understanding fluid-structure interactions near boundaries in biological systems.
\end{abstract}

\maketitle


\section{Introduction }

Locomotion and fluid pumping near surfaces are ubiquitous phenomena in nature, observed across a wide range of scales and fluid environments. In the aerial regime, birds such as swallows and seagulls frequently fly close to water or ground surfaces to hunt, presumably with the aerodynamic advantages provided by the ground effect \cite{Johansson2018, Blake1983, Webb2002}. Similarly, bats perform complex maneuvers just above ponds to drink water on the wing, a behavior that requires precise altitude control and lift force generation \cite{Maitra2025, Greif2010}. In aquatic environments, organisms range from dolphins utilizing the free surface to move \cite{Fish1991} to microscopic larvae suspending themselves at the interface. These behaviors are often driven by the need to access concentrated food resources, as seen in pond snails (\textit{Lymnaea stagnalis}) that filter-feed at the water-air boundary \cite{Pandey2023, Joo2020} or bivalves that pump nutrient-rich water near the sediment interface \cite{jorgensen1990bivalve}. Beyond resource acquisition, proximity to a boundary alters the fluid dynamics significantly, offering potential benefits such as drag reduction, lift enhancement, and improved propulsive efficiency \cite{Rayner1991, Zheng2024}.

However, the physical mechanisms governing these interactions vary drastically depending on the scale of the organism and the nature of the movement. While the aerodynamic ground effect for high-speed fliers is well-characterized by potential flow and vortex dynamics \cite{Mivehchi2021}, the hydrodynamics of slow-moving organisms near deformable interfaces remain less understood. For instance, freshwater snails utilize a muscular foot to generate localized currents \cite{Jung2021, Jung2025}, operating in a regime where viscosity dominates and the free surface can deform. Conversely, flapping fliers like bees and bats operate in inertial regimes where unsteady effects, such as vortex shedding and air compression, become significant.

In this manuscript, we present a comparative framework for animal interaction with surfaces, categorizing behaviors based on the Reynolds number ($Re$) and the Undulation number ($\text{Un}$). We contrast two distinct regimes: (1) Low Reynolds number undulation ($\text{Un} > 1$), exemplified by freshwater snails. We derive lubrication models to show how pumping and swimming speeds scale with the square of the amplitude-to-gap ratio $(a/h_0)^2$ and investigate the efficiency losses caused by surface deformation ($Ca/Bo \gg 1$). (2) High Reynolds number flapping ($\text{Un} < 1$), represented by bats and bees. We examine the ``drinking on the wing'' behavior in bats, revealing a 2.5-fold lift enhancement driven by the aerodynamic squeezing effect. Finally, we analyze honeybee fanning, demonstrating how a jet-vortex mechanism utilizes the ground effect to transport pheromones efficiently against diffusion. These diverse examples illustrate how animals leverage distinct physical principles to master the ground effect for survival.

\begin{figure}[htbp]
    \centering
    \includegraphics[width=.95\textwidth]{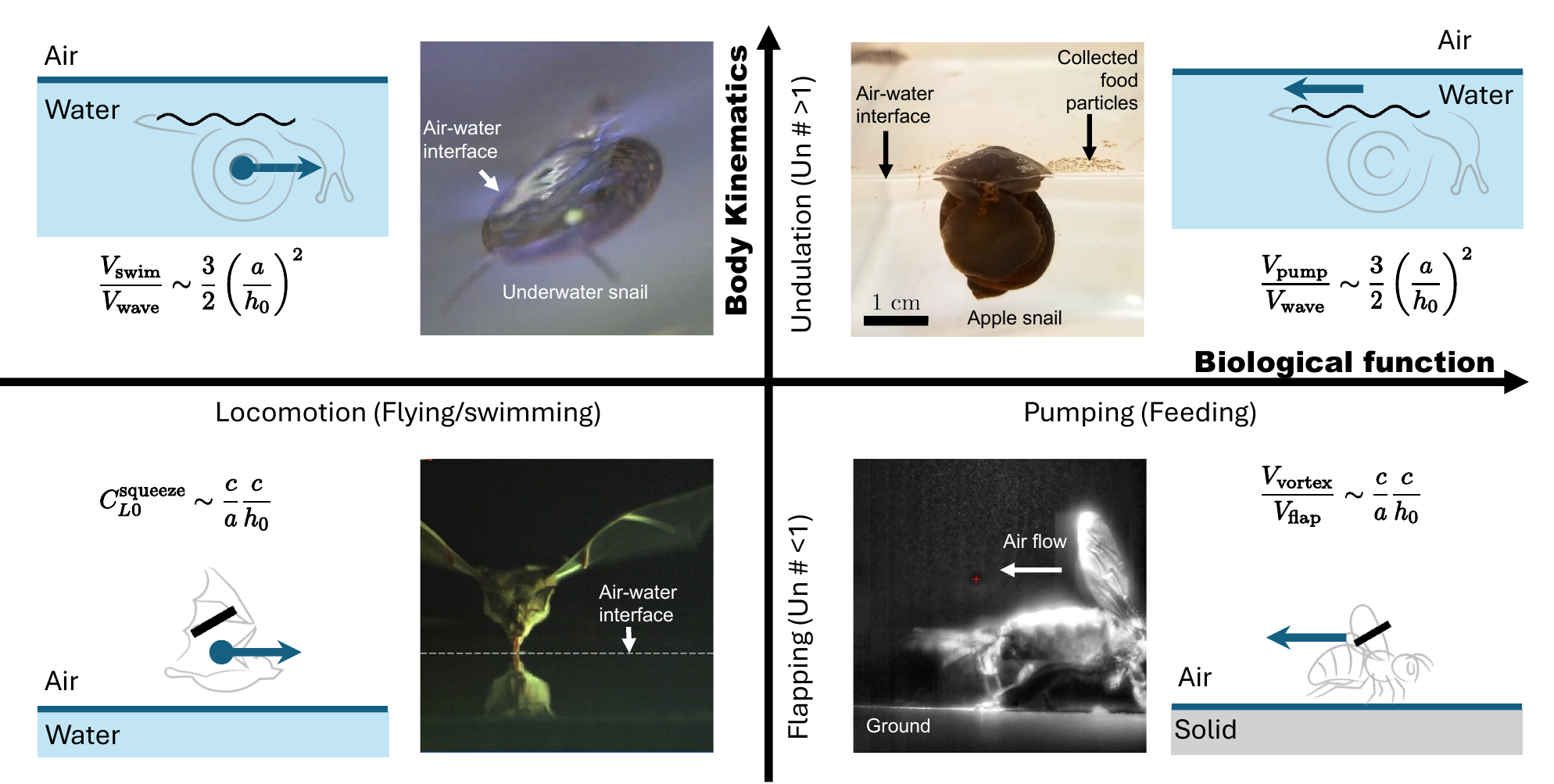}
    \caption{Comparative framework of animal locomotion and pumping near surfaces. The behaviors are categorized by the kinematic mode (Vertical axis: Undulation $\text{Un} > 1$ vs. Flapping $\text{Un} < 1$) and biological function (Horizontal axis: Locomotion vs. Pumping). (Top Row) Low Reynolds number undulation ($\text{Un} > 1$): Represented by freshwater snails near the air-water interface \cite{Pandey2023}. Whether swimming (left; credit: Drs. David Hu and Brian Chan) or stationary pumping for feeding (right; image from \cite{Joo2020}), the resulting velocity scales with the square of the wave amplitude relative to the gap height: $\frac{V}{V_{wave}} \sim \frac{3}{2} \left(\frac{a}{h_0}\right)^2$. (Bottom Row) High Reynolds number flapping ($\text{Un} < 1$): Represented by aerial fliers. (Left) A bat flying near the water surface ("drinking on the wing") experiences lift enhancement due to the squeezing effect, scaling with the chord-to-amplitude and chord-to-height ratios: $C_{L0}^{squeeze} \sim \frac{c}{a} \frac{c}{h_0}$ (right; image from \cite{Maitra2025}) A honeybee fanning near a solid ground generates a pumping flow (vortex street) that follows a similar geometric scaling: $\frac{V_{vortex}}{V_{flap}} \sim \frac{c}{a} \frac{c}{h_0}$ (left; image from \cite{Peters2017} and credit to Drs. Jake Peters and Stacey Combes).}
    \label{fig:overview}
\end{figure}

\section{Modes and Purpose of Undulation}

To classify animal movements or pumping, we utilize the Undulation number ($\text{Un} $), defined as the ratio of the active body length ($L_{active}$) to the undulation wavelength ($\lambda$). This dimensionless number provides a continuum for categorizing propulsion/pumping strategies across different fluid media. Biological observations reveal a wide range of animals that undulate their entire bodies and those that rely on oscillating appendages \cite{Lindsey1978}. In the aquatic realm, undulatory modes are traditionally classified by the fraction of the body participating in the wave motion. At one end of the spectrum lies the Anguilliform mode ($\text{Un} \approx 1.0$), exemplified by eels and lampreys \cite{Tytell2004}. In this mode, the entire body undulates, with more than one full wave fitting along the body length ($\text{Un} > 1$). While highly efficient in viscous or granular media, this whole-body undulation becomes less efficient at high swimming speeds. As we move towards faster swimmers, the undulation becomes increasingly confined to the posterior. In the Thunniform mode ($\text{Un} \approx 0.5$), seen in tuna and sharks, movement is concentrated almost entirely in the caudal fin. Here, the tail acts as a rigid hydrofoil pivoting to generate thrust/lift, offering high efficiency for sustained high-speed swimming \cite{Bainbridge1963}. The extreme case is the Ostraciiform mode ($\text{Un} \approx 0.25$), where propulsion is generated solely by the oscillation of a rigid tail fin, as seen in boxfish. 

In contrast to aquatic undulators, aerial fliers operate in the "Flapping" regime where the Undulation number is significantly less than unity ($\text{Un} < 1$). The mechanism here shift fundamentally from undulating body motions to lift generation via oscillating airfoils. Experimental data highlights this trend across different scales of flight. Insects, employing high-frequency flapping with rigid wings, operate at an Undulation number of approximately $0.24 \pm 0.12$. Bats, which utilize flexible airfoils capable of active cambering, show even lower values around $0.11 \pm 0.06$ (based on the data in \cite{Jung2021}). Birds exhibit the lowest undulation numbers of all, at approximately $0.06 \pm 0.02$, relying on semi-rigid aerofoils to generate lift. Despite these differences in morphology, efficient cruising across these groups is characterized by a Strouhal number ($St = fA/V_{fly}$) generally falling between 0.2 and 0.4 \cite{taylor2003flying}. 

Beyond kinematics, the interaction between the animal and the fluid serves two distinct biological functions: locomotion and pumping. Locomotion is defined as momentum transport, where the primary goal is thrust generation to transport the body through a stationary fluid. In this case, the body moves while the net fluid motion is minimized, as seen in fish swimming or birds flying. Conversely, pumping represents mass transport, where the body remains fixed (or effectively stationary) and the goal is to transport fluid past the structure. This function is critical for physiological processes such as feeding, respiration, and circulation, exemplified by the heart pumping blood or mussels filtering water \cite{Vogel2013}.

\begin{table}[ht]
\centering
\caption{Comparison of Undulation Mechanics across Species \cite{Jung2021, Lindsey1978, Tytell2004, Bainbridge1963, Lai2010}}
\begin{tabular}{|l|c|c|c|l|}
\hline
\textbf{Animal} & \textbf{Mode} & \text{Un} ($L_{active}/\lambda$) & \textbf{Mechanics} \\
\hline
Snail & Crawling & $> 2.0$ & Friction (Lubrication). Multiple waves fit on body \cite{Lai2010, Denny1980}. \\
Eel & Undulatory & $\sim 1.2$ & Reactive Fluid Force. One full wave fits on body \cite{Tytell2004}. \\
Trout & Undulatory & $\sim 0.6$ & Mixed. Half a wave fits on active tail \cite{Bainbridge1963}. \\
Tuna & Oscillatory & $0.2-0.4$ & Lift. Tail is a rigid hydrofoil that pivots \cite{Lindsey1978}. \\
Bird/Bat & Flapping & $0.2-0.4$ & Lift. Wing is a rigid/flexible aerofoil \cite{Jung2021}. \\
Insect & Flapping & $\sim 0.3$ & Lift. Wing is a rigid aerofoil. \\
\hline
\end{tabular}
\label{tab:mechanics}
\end{table}

\section{Non-dimensional numbers} 

To understand the dynamics of animals moving near surfaces, it is helpful to introduce several key non-dimensional numbers that characterize the interplay between inertial, gravitational, surface tension, and viscous forces in these environments. The Froude number ($Fr$) compares inertial forces to gravitational forces and is defined as \( Fr = {U}/{\sqrt{gL}} \), where $U$ is the characteristic velocity, $g$ is gravitational acceleration, and $L$ is a characteristic length. This number is particularly relevant for swimming and flying animals near surfaces, as it helps describe wave generation and locomotion efficiency. The Weber number ($We$) quantifies the ratio of inertial to surface tension forces and is given by \( We = {\rho U^2 L}/{\sigma} \), where $\rho$ is fluid density, $U$ is velocity, $L$ is characteristic length, and $\sigma$ is surface tension. This is important for animals interacting with water surfaces, such as water striders or flying fish, where surface tension plays a significant role. The Reynolds number ($Re$), defined as \( Re = {\rho U L}/{\mu} \) with $\mu$ as dynamic viscosity, expresses the ratio of inertial to viscous forces. It helps determine the flow regime, whether laminar or turbulent, around swimming or flying animals. The Bond number (Bo) compares gravitational and surface tension forces, given by \( Bo = {\rho g L^2}/{\sigma} \). This number is useful when considering the deformation of surfaces under the weight of an organism, as seen in insects standing on water. Finally, the Capillary number (Ca) measures the relative effect of viscous forces versus surface tension, defined as \( Ca = {\mu U}/{\sigma} \), and is relevant for small-scale locomotion and fluid pumping near interfaces. Together, these non-dimensional numbers provide a framework for analyzing and comparing the physical principles governing animal movement and fluid manipulation near surfaces in both aerial and aquatic environments.

\section{Locomotion/pumping in animals near a free surface}

\subsection{Snail locomotion}

We analyze the locomotion of a snail moving underwater near a free surface (air-water interface) \cite{Pandey2023, Joo2020}. The snail moves underneath the water surface with the average thickness $h_0$. The foot generates a traveling wave with wavelength $\lambda$, amplitude $a$, and wave speed $V_{wave}$. This dynamics may differ from crawling on a solid substrate because the upper boundary is a stress-free interface rather than a no-slip wall \cite{Lee2008, Chan2005}.

\begin{figure}[htbp]
    \centering
    \includegraphics[width=.9\textwidth]{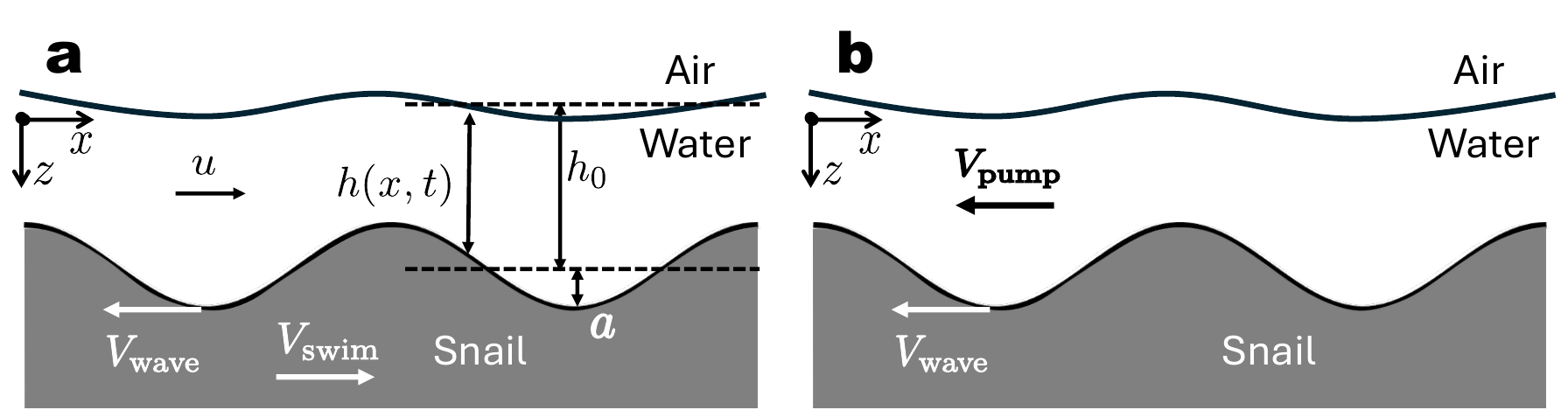}
    \caption{Schematic model of snail locomotion and pumping near a free surface.(a) Locomotion: A snail swims with velocity $V_{swim}$ near an air-water interface \cite{Pandey2023}. The snail's foot generates a traveling wave with propagation speed $V_{wave}$ and amplitude $a$. The gap height $h(x,t)$ oscillates around a mean film thickness $h_0$. The free surface (air-water interface) is modeled as a stress-free boundary that deforms in response to the pressure field. (b) Pumping: In the stationary pumping mode (where the snail is fixed), the same traveling wave motion ($V_{wave}$) generates a net fluid flow, $V_{pump}$, in the direction opposite to the wave propagation. This mechanism is used by snails for filter-feeding near the water surface.}
    \label{fig:snail}
\end{figure}

\begin{figure}[htbp]
    \centering
    \includegraphics[width=.9\textwidth]{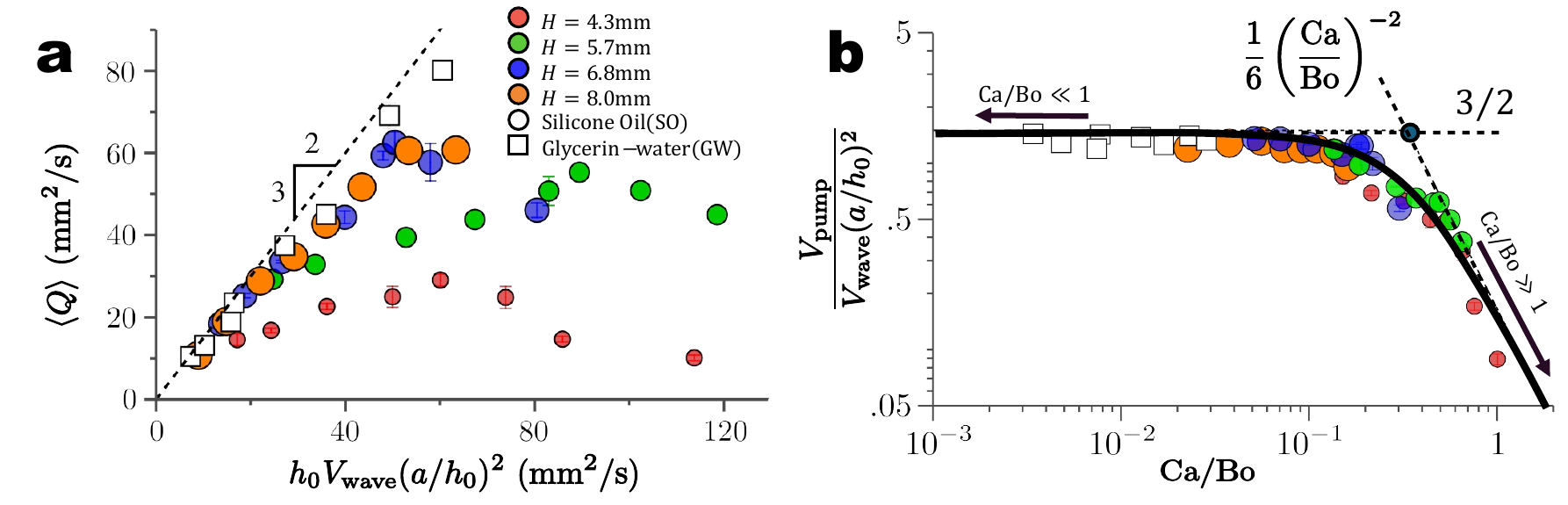}
    \caption{Experimental validation of snail pumping efficiency and the effect of surface deformation \cite{Pandey2023}.(a) Measured flow rate per unit width, $\langle Q \rangle$, plotted against the theoretical lubrication scaling parameter $h_0 V_{wave} (a/h_0)^2$. Symbols represent different gap heights ($H$) and working fluids (Silicone Oil and Glycerin-water). The deviation from the linear prediction (dashed line) at higher values indicates energy loss due to free surface interaction.(b) Normalized pumping efficiency $\frac{V_{pump}}{V_{wave}(a/h_0)^2}$ plotted against the ratio of the Capillary number to the Bond number ($Ca/Bo$). The plot collapses the data from (a) into two distinct regimes: Rigid Limit ($Ca/Bo \ll 1$): When gravity dominates, the surface remains flat, and the pumping efficiency plateaus at the theoretical limit of $3/2$. Deformable Limit ($Ca/Bo \gg 1$): When viscous forces dominate, the surface deforms significantly, reducing efficiency. The data follows a decay scaling of roughly $\frac{1}{6}(Ca/Bo)^{-2}$. }
    \label{fig:snail2}
\end{figure}

We assume the flow is governed by the lubrication approximation ($h_0/\lambda \ll 1$ and $Re \ll 1$). The coordinate system is defined such that $z=0$ represents the flat air-water interface, and $z=h(x,t)$ represents the undulating foot of the snail. The $x$-momentum equation for the thin film at low Reynolds number is given by the balance between the pressure gradient and viscous shear 
\begin{equation}
    \frac{\partial p}{\partial x} = \mu \frac{\partial^2 u}{\partial z^2} \,,
\end{equation}
where $p$ is the pressure, and $\mu$ is the fluid viscosity. The boundary conditions are: (1) At the air-water interface ($z=0$), the shear stress is zero, ${\partial u}/{\partial z} = 0$. (2) At the snail foot ($z=h$), the fluid moves with the wave speed, $u = -V_{wave}$. Integrating this momentum equation twice subject to these boundary conditions and an assumption of linear pressure along $x$-drection yields the velocity profile in the gap
\begin{equation}
    u(x,z) = \frac{1}{2\mu} \frac{dp}{dx} (z^2 - h^2) - V_{wave}\,.
\end{equation}
The swimming speed is determined by imposing a condition of zero net volume flux in the laboratory frame. A non-zero flux would imply fluid accumulation at one end, which contradicts the assumption of steady locomotion. For a snail swimming with velocity $V_{swim}$, the conservation of mass requires
\begin{equation}
    0 = \int_0^h u \, dz + h_0 (V_{swim} + V_{wave}) \,.
\end{equation}
Solving for the pressure gradient and applying the periodicity condition allows us to determine the swimming speed. For small amplitude undulations where $\epsilon = a/h_0 \ll 1$, the swimming speed scales as
\begin{equation}
    \frac{V_{swim}}{V_{wave}} \simeq -\frac{3 \epsilon^2}{2 + \epsilon^2} \approx -\frac{3}{2} \left( \frac{a}{h_0} \right)^2\,.
\end{equation}
This result highlights that locomotion near a stress-free boundary is driven by the square of the amplitude-to-gap ratio.

\subsection{Snail pumping}

In the absence of forward motion ($V_{swim} = 0$), the snail acts as a pump to transport fluid. We define the time-averaged flux per unit width, $\langle Q \rangle$, as the integral of the velocity profile over one period $\tau$ and the gap height 
\begin{equation}
    \langle Q \rangle = \frac{1}{\tau} \int_{0}^{\tau} \int_{0}^{h} u \, dz dt \simeq \frac{3}{2} h_0 V_{wave} \left(\frac{a}{h_0}\right)^2 \left( 1 + \mathcal{O}(Ca/Bo) \right) \,.
\end{equation}
The detailed calculation is in \cite{Pandey2023}. 
The averaged pumping speed $V_{pump}$ is then defined as the flux divided by the average gap height, $V_{pump} = \langle Q \rangle / h_0$. For a stiff interface where surface deformation effects are negligible ($Ca/Bo \ll 1$), this simplifies to the same scaling law derived for locomotion 
\begin{equation}
    \frac{V_{pump}}{V_{wave}} \simeq \frac{3}{2} \left( \frac{a}{h_0} \right)^2 \,.
\end{equation}

However, as the surface becomes deformable, the efficiency drops. Based on the experimental data shown in Figure~\ref{fig:snail2}, we incorporate a correction factor depending on the ratio of the Capillary number to the Bond number ($Ca/Bo$)
\begin{equation}
    \frac{V_{pump}}{V_{wave}(a/h_0)^2} \simeq \frac{3}{2} \left( \frac{1}{1 + 9(Ca/Bo)^2} \right)\,.
\end{equation}
This relation captures two distinct regimes: the rigid limit ($Ca/Bo \ll 1$) where the normalized speed plateaus at $3/2$, and the deformable limit ($Ca/Bo \gg 1$) where the pumping efficiency decays as approximately $(Ca/Bo)^{-2}/6$.

\subsection{Bat Flight}
We analyzed the flight of the Greater Horseshoe Bat (\textit{R. ferrumequinum}) and the Great Leaf-nosed Bat (\textit{H. pratti}) during water-surface drinking behaviors to understand the aerodynamic benefits of ground effect \cite{Maitra2025}. High-speed videography reveals distinct kinematic shifts when bats transition from straight flight to surface drinking. During drinking flights, the bat flies extremely close to the water surface ($h \approx 9.5$ cm for the small bat) to dip its mouth while keeping its wings dry. The kinematic comparison is summarized as follows: In straight flight, the bat operates with a pitch angle of $\approx 20^\circ$ and a large flapping amplitude of $\approx 6.6$ cm at a frequency of 8.7 Hz. In drinking flight, to avoid wing contact with the water, the flapping amplitude is drastically reduced to $\approx 2.4$ cm. To compensate for this reduction in stroke amplitude, the flapping frequency increases to 14.2 Hz. To quantify the aerodynamic forces, we modeled the drag and lift coefficients ($C_D$ and $C_L$) based on the angle of attack $\alpha$. The drag coefficient is modeled as a shifted cosine function 
\begin{equation}
    C_D = 0.0086 + C_{D0} [1 - \cos(2\alpha)]\,. 
\end{equation}
The lift coefficient is modeled sinusoidally
\begin{equation}
    C_L = C_{L0} \sin(2\alpha) \,.
\end{equation}

\begin{figure}[htbp]
    \centering
    \includegraphics[width=.8\textwidth]{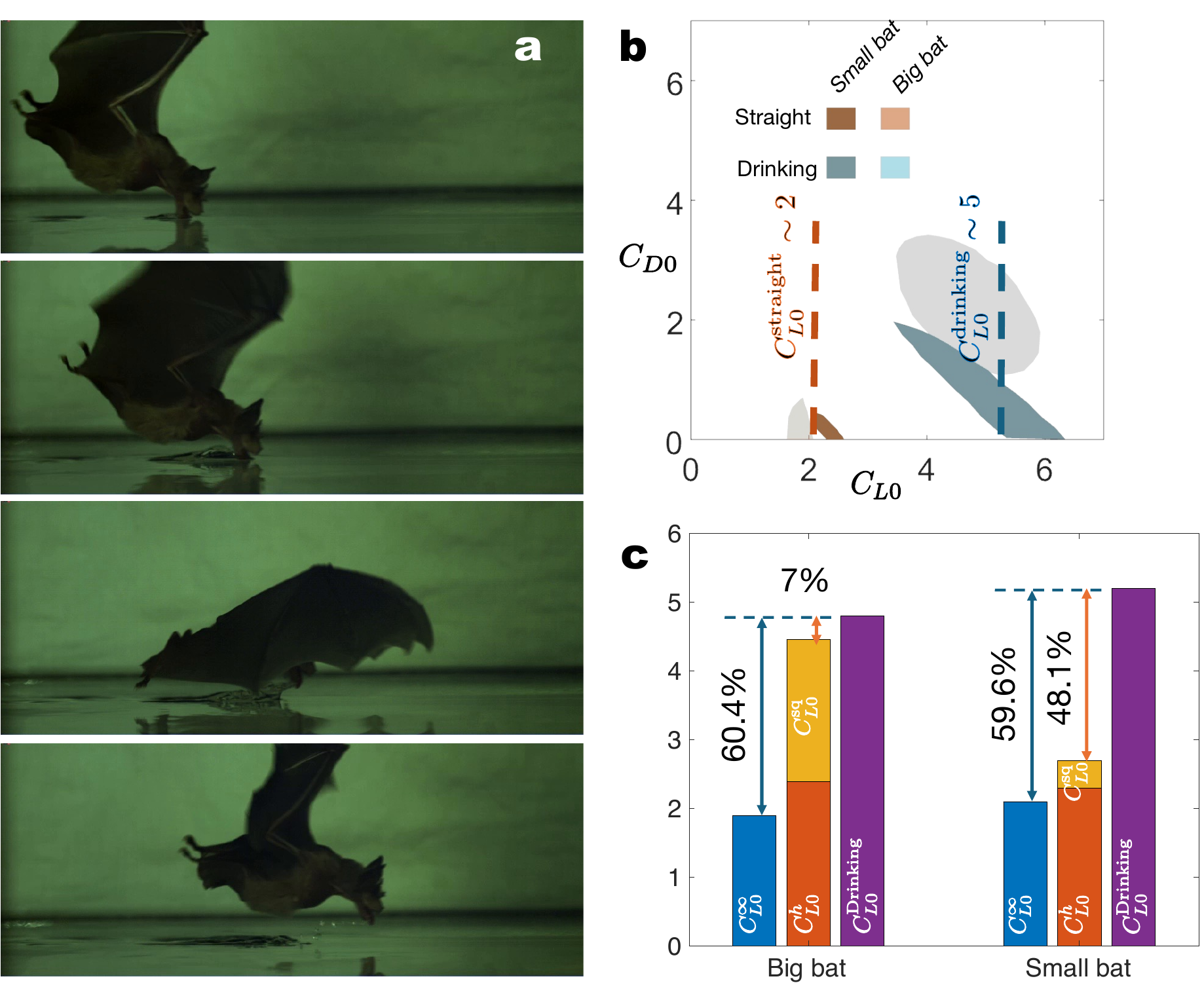}
    \caption{Aerodynamic coefficients and lift decomposition during bat drinking flight \cite{Maitra2025}. (a) A regime map plotting the drag parameter ($C_{D0}$) against the lift parameter ($C_{L0}$) for both the Big bat (H. pratti) and Small bat (R. ferrumequinum). The data reveals a distinct aerodynamic shift: during straight flight (orange/brown clusters), the lift parameter $C_{L0}^{straight}$ is approximately 2. However, during drinking flight (blue/teal clusters), $C_{L0}^{drinking}$ increases dramatically to approximately 5, representing a 2.5-fold enhancement. (b) Decomposition of the total lift coefficient during drinking flight ($C_{L0}^{Drinking}$). The observed lift is compared to the theoretical prediction from Weissinger’s potential flow model ($C_{L0}^h$, bottom bar segments). The potential flow model significantly underpredicts the total force. The remaining lift is attributed to the unsteady squeezing effect ($C_{L0}^{sq}$, top bar segments), which dominates the force generation, accounting for approximately 60\% of the total lift (60.4\% for the Big bat and 59.6\% for the Small bat).}
    \label{fig:bat2}
\end{figure}

Experimental data shows a dramatic increase in the lift parameter $C_{L0}$ during the drinking phase. For the small bat (\textit{R. ferrumequinum}), $C_{L0}$ increases from approximately 2 in open air (straight flight) to nearly 5 near the surface as shown in Figure \ref{fig:bat2}b.
\begin{equation}
    \frac{C_{L0}^{ground}}{C_{L0}^{\infty}} \approx 2.5
\end{equation}
This 2.5-fold increase allows the bat to support its weight despite the constraints (reduced amplitude) of the drinking maneuver. To determine if this lift enhancement can be explained by classical potential flow, we applied Weissinger's theory for a flapping foil near a boundary \cite{Weissinger1942}. This model utilizes the method of images, placing an image vortex with opposite circulation at depth $h$ below the air-water interface to satisfy the no-penetration condition.

\begin{figure}[htbp]
    \centering
    \includegraphics[width = .9\textwidth]{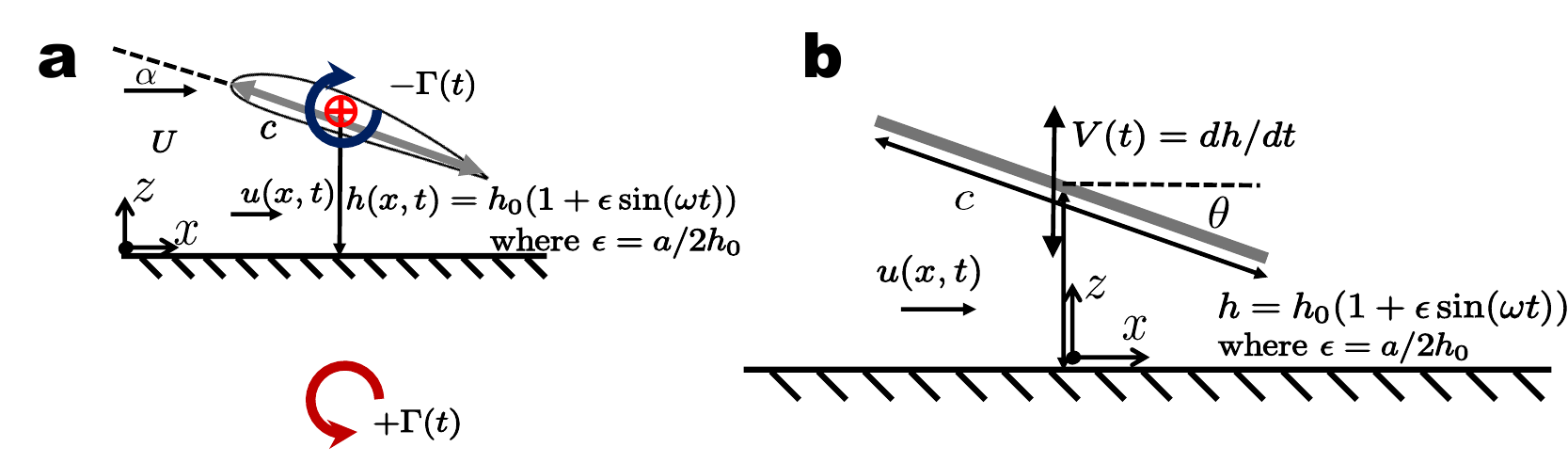}
    \caption{Theoretical models for aerodynamic ground effect in flapping flight.(a) Potential Flow (Method of Images): A discrete vortex model where the flapping wing  is represented by a bound vortex with circulation $-\Gamma(t)$. To satisfy the no-penetration boundary condition at the ground, an image vortex with opposite circulation $+\Gamma(t)$  is placed at an equal depth below the surface. The wing operates at a mean height $h_0$ with reduced frequency $\omega$. (b) Squeezing Flow (Air Cushion Model): A model for the unsteady pressure forces generated when the wing flaps close to the ground. The downward vertical velocity $V(t) = dh/dt$ compresses the air in the gap $h(t)$ , generating a high-velocity horizontal squeeze flow $u(x,t)$. This mechanism creates a pressure cushion that significantly enhances lift for small gap heights ($h \ll c$) and small pitch angles ($\theta$).}
    \label{fig:bat}
\end{figure}

The theoretical lift coefficient near the interface, $C_{L0}^h$, is derived as a function of the chord-to-height ratio ($c/h_0$) and the non-dimensional amplitude ($a/h_0$)
\begin{equation}
    C_{L0}^{h} = C_{L0}^{\infty} \left[ 1 + \frac{1}{16} \left(\frac{c}{h_0}\right)^2 \frac{1}{(1 - (a/h_0)^2)^{3/2}} \right]\,,
\end{equation}
where $C_{L0}^{\infty}$ is the lift coefficient in unbounded flow. Comparing the theoretical prediction with the experimental data reveals a significant discrepancy. Weissinger's potential flow model predicts only a modest increase in lift (approximately 12--25\%), leaving a large portion of the observed lift enhancement unexplained. Specifically, the theory fails to account for approximately 60.4\% of the lift for the big bat and 59.6\% for the small bat. This suggests that the dominant mechanism is not simple circulation enhancement but rather the "squeezing effect" (unsteady ground effect). When the wing flaps close to the ground, air is trapped and pressurized between the wing and the surface. We model the gap height $h(t)$ as a function of the mean height $h_0$ and the flapping amplitude $a$
\begin{equation}
    h(t) = h_0 (1 + \epsilon \sin(\omega t)) \quad \text{where} \quad \epsilon = a/2h_0\,.
\end{equation}
Assuming the horizontal flow is stagnant at the center ($x=0$) and the tilt angle is small ($\theta \ll 1$), the horizontal velocity profile $u(x,t)$ in the gap is governed by the squeeze rate $V(t) = dh/dt$
\begin{equation}
    u(x,t) = - \frac{V x}{h}\,.
\end{equation}
The pressure distribution is derived from the unsteady momentum equation
\begin{equation}
    -\frac{1}{\rho} \frac{\partial p}{\partial x} = \frac{\partial u}{\partial t} + u \frac{\partial u}{\partial x}\,.
\end{equation}
Integrating the pressure over the chord length $c$, the vertical force $F$ scales as
\begin{equation}
    F = \int_{-c/2}^{c/2} (p - p_0) dx \sim \rho \frac{c^3 s}{h} \left[ 1 + c_1 \left(\frac{c\theta}{h}\right)^2 \right] \frac{dV}{dt}\,.
\end{equation}
This allows us to derive a scaling law for the lift coefficient due to squeezing, $C_{L0}^{sq}$. Normalizing by the dynamic pressure, we find that the enhancement depends on two geometric ratios: the chord-to-amplitude ratio ($c/a$) and the chord-to-height ratio ($c/h_0$)
\begin{equation}
    C_{L0}^{sq} \propto \frac{c}{a} \frac{c}{h_0}\,.
\end{equation}
This air-cushion effect provides the necessary force augmentation for the ``drinking on the wing'' maneuver that potential flow models alone cannot capture.

\begin{table}[ht]
\centering
\caption{Morphological and Flight Parameters for \textit{H. pratti} (Bigbat) and \textit{R. ferrumequinum} (Smallbat) \cite{Maitra2025} }
\begin{tabular}{|l|c|c|c|c|c|c|}
\hline
{\textbf{Parameter}} & {\textbf{Sym}} & {\textbf{Units}} & \multicolumn{2}{c|}{\textbf{H. pratti (Bigbat)}} & \multicolumn{2}{c|}{\textbf{R. ferrumequinum (Smallbat)}} \\
\hline
Body mass & $m$ & g & \multicolumn{2}{c|}{$65 \pm 15$} & \multicolumn{2}{c|}{$17 \pm 3$} \\
Wing span & $b$ & cm & \multicolumn{2}{c|}{$56 \pm 1.1$} & \multicolumn{2}{c|}{$30.7 \pm 3.5$} \\
Chord length & $c$ & cm & \multicolumn{2}{c|}{$16.7 \pm 1$} & \multicolumn{2}{c|}{$9.1 \pm 0.8$} \\
Wing area & $A$ & cm$^2$ & \multicolumn{2}{c|}{$482 \pm 42$} & \multicolumn{2}{c|}{$146.4 \pm 17$} \\
\hline
\cline{4-7}
 & & & \textbf{Straight} & \textbf{Drinking} & \textbf{Straight} & \textbf{Drinking} \\
\hline
Pitch angle (mean) & — & deg & $28$ & $17$ & $20$ & $20.5$ \\
Mean distance from ground & $h$ & cm & — & $12 \pm 0.7$ & — & $9.5 \pm 3.5$ \\
Flapping amplitude & $\alpha$ & cm & $10.6 \pm 1.4$ & $6.7 \pm 1.3$ & $6.6 \pm 1.5$ & $2.4 \pm 0.4$ \\
Flapping frequency & $f$ & Hz & $9.8 \pm 0.4$ & $11.9 \pm 0.6$ & $8.7 \pm 1.2$ & $14.2 \pm 1.2$ \\
Flight speed & $U$ & m·s$^{-1}$ & $2.8 \pm 0.3$ & $1.6 \pm 0.2$ & $2.5 \pm 0.8$ & $1.1 \pm 0.2$ \\
\hline
\end{tabular}
\end{table}

\subsection{Honeybee Pump-fanning}
Honeybees (\textit{Apis mellifera}) utilize their wings not just for flight ($\text{Un} \ll 1$), but for stationary pumping to control their nest environment \cite{Peters2017}. They exhibit two distinct fanning behaviors with different kinematic signatures and flow outputs. First, ventilation fanning is used to cool the nest. Bees stand on the comb and fan at a lower frequency ($f \approx 174$ Hz) but with a large amplitude ($\approx 118^\circ$) to maximize bulk flow, achieving speeds up to $\sim 3.0$ m/s \cite{Peters2019}. Second, nasonov scenting is used to disperse pheromones to guide swarms. Bees raise their abdomens to expose the Nasonov gland and fan at a higher frequency ($f \approx 213$ Hz) with a smaller amplitude ($\approx 98^\circ$), generating a lower flow speed of $0.94 \pm 0.11$ m/s \cite{Nguyen2021}.

\begin{figure}[htbp]
    \centering
    \includegraphics[width = .7\textwidth]{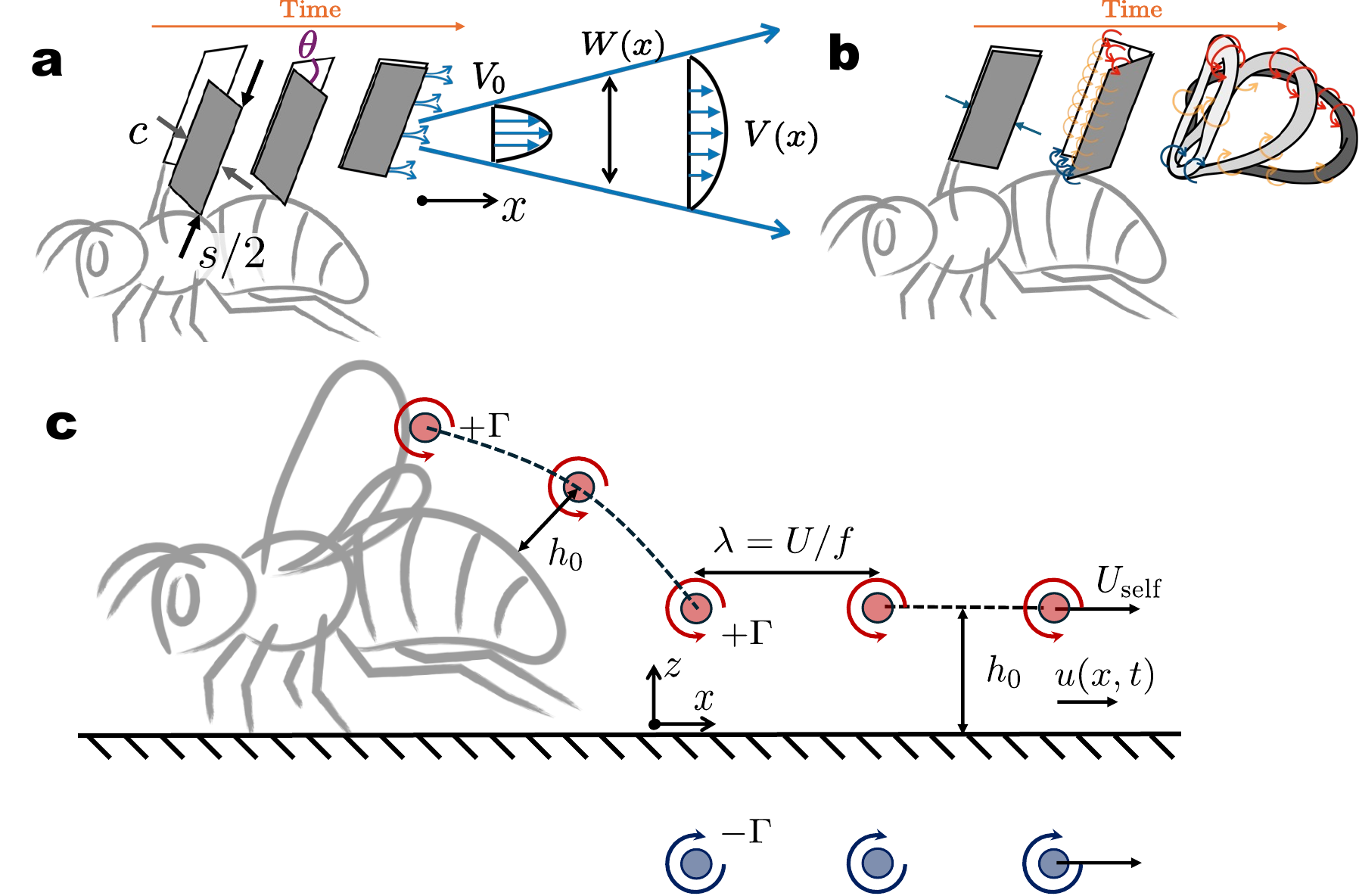}
    \caption{Schematic of the Jet-Vortex Pumping Mechanism in Honeybees. (a) The Clap Phase (Jet Formation): The wings close rapidly, acting as a piston to generate a high-momentum fluid jet with initial velocity $V_0$. As the jet propagates, it behaves as a turbulent jet where the centerline velocity $V(x)$ decays inversely with distance $x$, and the jet width $W(x)$ expands, leading to rapid dispersion of the fluid. (b) The Fling Phase (Vortex Generation): The wings peel apart, generating a coherent vortex ring with circulation $\Gamma$ at the wingtips. The lower diagram illustrates the theoretical model of a vortex dipole near a solid boundary ($z=0$) using the method of images. A real vortex ($+\Gamma$) located at height $h_0$ interacts with its image vortex ($-\Gamma$) at depth $-h_0$, creating a self-induced velocity $U_{self}$. Unlike the dispersing jet, these coherent vortices encapsulate the pheromones, allowing them to be advected downstream with minimal dilution \cite{Peters2017}.}
    \label{fig:bee}
\end{figure}

The pumping mechanism relies on a clap-and-fling motion that generates a hybrid flow structure consisting of a high-momentum jet and coherent vortex rings \cite{WeisFogh1973}. 

During the initial clap phase, the wings close rapidly to generate a high-momentum jet. The initial puffing air speed at the wings, $V_0$, is determined by the volume change relative to the opening area, scaling as $V_0 \sim 2\pi f c \approx 4.3$ m/s. While this velocity is high, a simple turbulent jet is inefficient for long-range chemical signaling due to rapid decay. For a 3D turbulent jet, the centerline velocity $V(x)$ decays inversely with distance according to the scaling law
\begin{equation}
    \frac{V(x)}{V_0} \approx K \frac{c}{x}\,,
\end{equation}
where $K \approx 5$ is a decay parameter established in turbulent flow literature \cite{Pope2000, Tennekes1972}. Based on this relation, the velocity drops to 10\% of its initial value ($0.1 V_0$) at a distance of approximately $x \approx 15$ cm. Furthermore, momentum balance implies that the jet width $W$ increases significantly with distance:
\begin{equation}
    c \cdot V_0 = W \cdot V(x) \implies W = \frac{x}{K}
\end{equation}
At a distance of $x=15$ cm, the jet width expands to $W \approx 3$ cm. This significant expansion results in a pheromone signal that is too dispersed to be effective for directional communication. 

Following the clap, the bee utilizes the fling phase to mitigate the rapid dispersion of the jet. The rapid wing opening generates a vortex ring with circulation $\Gamma$ scaling as
\begin{equation}
    \Gamma \simeq c V_0 = 2\pi f c^2 \approx 0.013 \, \text{m}^2/\text{s}\,.
\end{equation}
The self-induced velocity $V_{vortex}$ of this vortex ring near a wall is given by the method of images \cite{Lamb1932}
\begin{equation}
    V_{vortex} = \frac{\Gamma}{2\lambda} \coth\left(\frac{2\pi h_0}{\lambda}\right) \approx \frac{\Gamma}{4\pi h_0} = \frac{c^2 f}{2 h_0} \approx 0.45 \, \text{m/s} \,.
\end{equation}
Notably, this self-induced speed is significantly lower than the initial jet velocity ($0.45 \, \text{m/s} \ll 4.3 \, \text{m/s}$). Consequently, the vortices do not primarily propel themselves; rather, they are advected by the strong jet. The crucial function of the vortex structure is to encapsulate the pheromones, preventing the rapid mixing and dilution that occurs in a standard turbulent jet. The lifespan of these vortices is governed by viscous dissipation. The dissipation time scale $t_{diss}$ is
\begin{equation}
    t_{diss} \approx \frac{h_0^2}{4\nu_{air}} \approx 0.09 \, \text{s} \,.
\end{equation}
This duration is orders of magnitude longer than the flapping period ($1/f \approx 0.005$ s), allowing the structure to persist. By integrating the decaying jet velocity $V(x)$ over the vortex lifespan ($dt = dx/V(x)$), we derive the maximum effective propagation distance $X_{max}$
\begin{equation}
    X_{max} = \sqrt{2 \, V_0 K c \, t_{diss}} \sim 10.7 \, \text{cm}\,.
\end{equation}
This distance of $\sim 10.7$ cm exceeds the typical spacing between scenting bees in a swarm (5--7 cm). Thus, the "Jet-Vortex" mechanism ensures that pheromone-laden vortices retain sufficient momentum and coherence to bridge the gap to the next bee, allowing the signal to be detected and re-amplified \cite{Nguyen2021}.

\section{Conclusion}

This study establishes a unified physical framework for understanding how animals utilize proximity to surfaces for locomotion and fluid transport. By categorizing these behaviors based on the Undulation number ($\text{Un}$) and Reynolds number ($Re$), we distinguish between two fundamental regimes: viscous-dominated undulation ($\text{Un} > 1$, low $Re$) and inertia-dominated flapping ($\text{Un} < 1$, high $Re$).

For low Reynolds number swimmers like freshwater snails, locomotion and pumping are governed by lubrication forces because the fluid gap is much smaller than the undulation wavelength. Our theoretical models and robotic experiments confirm that swimming and pumping speeds scale with the square of the amplitude-to-gap ratio, $(a/h_0)^2$. Furthermore, we identified a critical performance limit imposed by the free surface; when the surface is deformable (high Capillary-to-Bond number ratio, $Ca/Bo \gg 1$), energy is used and dissipated into surface deformation rather than fluid transport, significantly reducing efficiency.

In the high Reynolds number regime, aerial fliers exploit ground effects to enhance aerodynamic forces. For bats drinking on the wing, we observed a massive 2.5-fold increase in the lift coefficient constant ($C_{L0} \approx 5$) compared to open-air flight. This enhancement cannot be fully explained by potential flow theory (method of images) alone but is largely driven by the unsteady squeezing effect, which scales with the chord-to-height ratio ($c/h_0$).
Similarly, honeybees utilize ground proximity for chemical communication via a jet-vortex pumping mechanism. By shifting their kinematics to a high-amplitude clap-and-fling mode, bees generate coherent vortex rings. Unlike simple turbulent jets that decay rapidly, these vortices encapsulate pheromones and transport them over distances of $\sim 10$ cm, effectively bridging the gap between individuals in a swarm against diffusive losses.

These findings demonstrate that the ground effect is a vital ecological benefit to many organisms. 



\section{Acknowledgement}
The author gratefully acknowledges Dr. Abhradeep Maitra for providing bat flight data, and Dr. Kirstin Peterson and Ms. Lydia Calderon for valuable discussions on bee pump-fanning behaviors. Additionally, the author acknowledges the use of AI tools for proofreading and language refinement during the preparation of this manuscript.

\section{Funding}
This work is supported by NOAA Sea Grant NA24OARX417C0598-T1-01.

\section{Competing interests}
The author declares no competing financial interest.

\section{References}


\end{document}